\documentstyle[12pt]{article}
\def\rmdj {d\llap{\raise 1.22ex\hbox
  {\vrule height 0.09ex width 0.315em}\kern 0.04em}}
\def\sldj {d\llap{\raise 1.22ex\hbox
  {\vrule height 0.09ex width 0.265em}}\rlap{\raise 1.22ex\hbox
  {\vrule height 0.09ex width 0.05em}}}
\def\itdj {d\llap{\raise 1.22ex\hbox
  {\vrule height 0.09ex width 0.2em}}\rlap{\raise 1.22ex\hbox
  {\vrule height 0.09ex width 0.06em}}}
\def\bfdj {d\llap{\raise 1.16ex\hbox
  {\vrule height 0.126ex width 0.308em}\kern 0.04em}}
\def\ttdj {\rlap{\kern 0.17em\raise 1.1ex\hbox
  {\vrule height 0.09ex width 0.295em}}d}
\def\scdj {\rlap{\kern 0.04em\raise 0.57ex\hbox
  {\vrule height 0.09ex width 0.20em}}d}
\def\sfdj {d\llap{\raise 1.22ex\hbox
  {\vrule height 0.10ex width 0.3em}\kern 0.02em}}

\def\dj{\ifcase\fam \rmdj \or \or \or
  \or \itdj \or \sldj \or \bfdj \or \ttdj \or \sfdj \or \scdj
    \else \rmdj \fi}

\newcommand{\lessoe}{\raisebox{0.25ex}{$<$}\hspace*{-0.78em}
      \raisebox{-0.95ex}{$\sim$}}
\newcommand{\beq}{\begin{equation}}
\newcommand{\eeq}{\end{equation}}

\begin{document}

\thispagestyle{empty}   

\begin{flushleft}
BI--TP 94/21 \\
ZTF - 94/05\\
\end{flushleft}

\vspace*{2.0cm}

\begin{center}

  \begin{Large}

  \begin{bf}

Quark off-shell contributions to $K_{L}\rightarrow\gamma\gamma$\\
in a bound-state approach

  \end{bf}

  \end{Large}

\vspace{1cm}

  \begin{Large}

 D. Kekez\footnotemark,
D. Klabu\v{c}ar\footnotemark\addtocounter{footnote}{-1},
K. Kumeri\v{c}ki\footnotemark\addtocounter{footnote}{-1} and
I. Picek\footnotemark\addtocounter{footnote}{-2}
\vspace{1.0cm}

$^{1}${\small Ru\dj er Bo\v{s}kovi\'{c} Institute,
POB 1016, 41001 Zagreb, Croatia\\}

$^{2}${\small Department of Physics, Faculty of Science, University
of Zagreb,
POB 162,\\ HR-41000 Zagreb, Croatia\\}

\end{Large}

\end{center}

\vspace{0.7cm}

\begin{center}
  {\bf Abstract}
\end{center}
\begin{quotation}
\noindent
   We present a new piece of evidence in favour of the importance
of the quark off-shellness in the kaon.
The matrix elements of the
flavour-changing operators for $K_{L}\rightarrow\gamma\gamma$
comply with the general behaviour of the matrix elements expected
in the pairing bound-state model used here.
The present calculation in essence agrees
with previous chiral-quark results.
The off-shell contribution
turns out to be dominant (the model on-shell
amplitude being at the 10\% level).
Compared with the chiral perturbation-theory approach,
our off-shell contribution is an entirely new  ${\cal{O}}(p^{4})$
{\em direct-decay} piece, whereas the non-diagonal magnetic-moment
term belongs to the order ${\cal{O}}(p^{6})$.

\end{quotation}

\newpage

\section{Introduction}

   There is a long-standing need to calculate hadronic matrix
elements properly -- a problem that involves departure from the parton
(free-quark mass-shell) regime. A recent attempt to partially
fulfill this programme has focused on the study of quark off-shell
effects in radiative $K^{0}$-decays \cite{ep93,ep94}.
These off-shell contributions were revealed by Eeg and Picek
first in the significant CP-violating $K_{L}\rightarrow\gamma\gamma$
amplitude \cite{ep93} and more recently in the ``anomalous part'' of
the CP-conserving $K_{L}\rightarrow\gamma\gamma$ amplitude
\cite{ep94}. The theoretical framework in which off-shell
quarks were handled was an effective low-energy QCD model \cite{chqm},
the ``chiral quark model" providing the meson-quark coupling.
Such a framework departs from the perturbative (partonic) QCD regime,
where one faces the vanishing of the matrix elements of those operators
\cite{polsim} which are zero if QCD equations of motion (EOM) are used.
However, presently it is
possible to go beyond the perturbative analysis of Politzer and Simma
\cite{polsim}  only
in the model dependent way. Therefore the off-shell effect calculated
in the chiral quark model should be studied in other approaches
accounting for the nonperturbative QCD.

The purpose of this paper is to investigate off-shell effects in a
completely different type of model where mesons are not treated as
elementary fields, but are represented by quark-antiquark bound states.
Such a model is a natural environment for studying off-shell
 effects, as quarks are by
definition off-shell in the bound states, and especially
so in such strongly bound, highly relativistic systems as
light pseudoscalar mesons.
Another advantage is that bound-state solutions are not pointlike
but extended, and this will by itself ensure that the quark
loops in our calculations do not diverge. As has often been  pointed out
(e. g. recently
by Ball and Ripka \cite{ball}), {\it ad hoc} regularization procedures
often lead to inconsistencies and spurious results.
In the present approach, no {\it ad hoc}
regularizations or cut-offs are necessary,
because the momentum dependence of the non-local bound-state
meson vertices provides a  natural regularization.

The calculation of the matrix elements of relevant quark
operators will proceed along the same track as the previous calculation
of pion and kaon decay constants and the $\pi^{0}\rightarrow\gamma\gamma$
amplitude in the quark bilocal bound-state model \cite{hkkp}. Using
$\pi^{0}\rightarrow\gamma\gamma$ as a ``monitoring process'' has the
advantage of decoding the anomaly part in $K_{L}\rightarrow\gamma\gamma$
in the same way in which the comparative consideration of these
processes
in variants of the effective low-energy QCD enabled us to isolate the
anomalous contribution in \cite{ep94}.

As in  \cite{ep94}, let us again parametrize the
$K_{L}\rightarrow\gamma\gamma$ and
$\pi^{0}\rightarrow\gamma\gamma$ processes by an effective interaction
of order ${\cal{O}}(\alpha=e^2/4\pi)$:
\beq
{\cal L}_P \, = \,
 { \alpha  C_P} \epsilon_{\mu\nu\rho\sigma}
F^{\mu\nu}F^{\rho\sigma} \phi_{P}\; ,
\label{eq:lagp}
\eeq
where,
for $P=K_{L}\simeq K_{2}=\frac{1}{
\sqrt{2}}(K^{0}+\bar{K}^{0}) $ or $P=\pi^{0}$,
the measured widths 
require
\begin{center}
  $|C_{K_{2}}|=5.9\times 10^{-11} \mbox{MeV}^{-1}\;\;\;\; ;
   \;\;\;\;|C_{\pi^{0}}|=4.3\times 10^{-4} \mbox{MeV}^{-1}\qquad .$
\end{center}
The low-energy QCD calculation of ref. \cite{ep94} accounted  for the
full $C_{\pi^{0}}$ (axial  anomaly) amplitude.
In this reference the authors were able
to isolate the anomalous part of the $K_{L}\rightarrow\gamma\gamma$
amplitude and showed that it accounted for roughly a quarter\footnote{
Instead of the half quoted in ref. \cite{ep94}. There was a
superfluous factor of 2 in eq. (8) of this reference.} of the
empirical $|C_{K_{2}}|$. This amplitude, resembling very much
the famous anomalous  pionic one, is  essentially a
{\em direct} contribution, as opposed to possible {\em reducible} pole
contributions, which is out of scope of both the
previous \cite{ep93,ep94} and the present work.
Actually, the anomalous part of the
$K_{L}\rightarrow\gamma\gamma$ amplitude was represented by the
off-shell contribution from the point of view of the effective
quark-operator evaluation.

  Let us now recall the appearance of the off-shellness \cite{ep93,ep94}
 in $K_{L}\rightarrow\gamma\gamma$.
The essential point is to overbridge the nonperturbative QCD
($\lessoe$ 1 GeV)
and the electroweak ($\sim M_{W}$) scale. At the latter, the flavour
change (FC) $s\rightarrow d$ in the presence of external photons
results (after integrating out heavy loop particles) in an effective
lagrangian \cite{ep93}
\beq
{\cal{L}}(s\rightarrow d)_{\gamma} \, = \,  B \,
\epsilon^{\mu \nu \lambda \rho}
F_{\mu \nu} \, ( \bar{d} \; i \stackrel{\leftrightarrow}{D_{\lambda}}
 \gamma_{\rho} L\, s )  \; ,
\label{eq:lagg}
\eeq
where quarks are interacting fields with respect to QCD. This fundamental
fields are chiral fermions ($L=\frac{1-\gamma_{5}}{2}$,
 $R=\frac{1+\gamma_{5}}{2}$ projections). As explained
previously \cite{ep93}, dealing with free quarks at such a high-energy
scale results in cancellation of 1PI and 1PR graphs for
$s\rightarrow d\gamma\gamma$, induced by $V_{1\gamma}$ and $V_{
2\gamma}$ vertices contained in (\ref{eq:lagg}) and explicated in
eqs. (\ref{eq:v1}) and (\ref{eq:v2}) below. However, when bringing
these vertices
below the GeV scale, they operate on the off-shell (bound) quark
states and the cancellation can be lost. In order to follow the
fate of the off-shell contribution, it is convenient to rewrite
(\ref{eq:lagg}) in the form
\beq
{\cal{L}}(s\rightarrow d)_{\gamma} = {\cal{L}}_F +
{\cal{L}}_{\sigma} \qquad ,
\label{eq:lagfs}
\eeq
where
\beq
{\cal{L}}_F = B \, \bar{d} [(i\gamma \cdot D - m_d) \,
 \sigma_{\mu \nu} F^{\mu \nu} L + \sigma_{\mu \nu} F^{\mu \nu} R (i\gamma
\cdot D -m_s)] s
\label{eq:lagf}
\eeq
represents the piece which would vanish on-shell by applying the
QCD equations of motion, and
\beq
{\cal{L}}_{\sigma} = B \, \bar{d} \, (m_s \sigma_{\mu \nu} F^{\mu \nu} R +
 m_d  \sigma_{\mu \nu} F^{\mu \nu} L) \, s
\label{eq:lags}
\eeq
is the off-diagonal magnetic-moment term which vanishes in the
chiral limit. The quantity $B\sim eG_{F}$,
\beq
   B=\frac{G_{F}}{\sqrt{2}}\frac{e}{4\pi^{2}}\lambda_{u}
    \hat{B}_{u}\qquad ,
\label{eq:defb}
\eeq
contains the Kobayashi--Maskawa factor
of the value $\lambda_{u}=V_{ud}V_{us}^{*}\simeq 0.215$, whereas
$\hat{B}_{u}$ incorporates the perturbative, short-distance QCD
corrections.
The precise value of this $\hat{B}_{u}$ factor depends on the
renormalization scale $\mu$ \cite{svz}, increasing from
0.16 for $\mu = $0.7 GeV to 0.66 for $\mu=$0.3 GeV for the local
operator in (\ref{eq:lagf}), and ranging between
0.14 for $\mu = $0.7 GeV and 0.32 for $\mu=$0.3 GeV for the local
operator of the nondiagonal magnetic-moment transition in
(\ref{eq:lags}).
Somewhat different behaviour of these two operators at
lower-energy scales is due to the different anomalous dimensions they
have.

In the present paper  the calculation in the
chiral limit suffices to extract the chiral-anomaly contribution
(which is known to be an overwhelming contribution to the
$\pi^{0}\rightarrow\gamma\gamma$ decay).
However, the kaon decay requires investigation
beyond the chiral limit. Therefore,
in the following we first present the calculation in the
$SU(3)_f$ limit ($m_q = m_{u,d} = m_s$), suitable for
extracting the anomalous contribution in the chiral limit. After that
we proceed with the study of $K_{L}\rightarrow\gamma\gamma$ beyond the
chiral limit and for non-degenerate quark masses. This enables us
to find a share of the non-anomalous part in the direct
$K_{L}\rightarrow\gamma\gamma$ decay amplitude.

\section{Bound-state evaluation of the $K_{L}\rightarrow\gamma\gamma$
         amplitude}
To evaluate the hadronic matrix elements  of the above effective
operators (\ref{eq:lagg})--(\ref{eq:defb}),
 we use the variant \cite{pervPL,perv} of an effective meson bilocal
theory [9--12] in which
the related decay $\pi^{0}\rightarrow\gamma\gamma$ was computed by two
of us \cite{hkkp}, along with the pion and kaon spectrum
and the decay constants\footnote{$F_{K}=\sqrt{2}f_{K}$, where
 $f_{K}=f_{\pi}$ ($f_{\pi}^{expt}=92.4 MeV$) in the $SU(3)_f$ symmetry
limit.}
$F_{\pi}$ and $F_{K}$.

Varying the effective bilocal action yields
[9--12, 7] the Schwinger--Dyson equation
(SDE) for the dressed quark propagator (whose self-mass $\Sigma(q)$
is thereby generated dynamically), and the Bethe-Salpeter equation
(BSE) for the bilocal bound-state meson-vertex function.  The
meson mass $M_P$ results as the corresponding BSE eigenvalue.

Since we are interested in the qualitative issue of the existence
and the importance of off-shell effects and not in the precise
quantitative description of hadrons as bound states, we choose a very
simplified instantaneous quark-quark interaction kernel $K(x,y)$ leading
to a potential model with very tractable SDE and BSE. Concretely,
we use the special form \cite{pervPL,perv}
\beq
   K^{\eta}(x-y)=K^{\eta}(z,X)=\eta_{\mu}\gamma^{\mu}V(z_{\perp})
   \delta(z_{P})\eta_{\nu}\gamma^{\nu} \qquad ,
\label{eq:defk}
\eeq
where $z=x-y$, $X=(x+y)/2$ and $\eta^{\mu}=P^{\mu}/\sqrt{P^{2}}$,
whereas the decomposition of four-vectors into  components parallel
and perpendicular to the total meson momentum $P^{\mu}$ is given
by $x^{\mu}_{\parallel}=\eta^{\mu}x_{P}$, $x_{P}=x\cdot\eta$ and
$x^{\mu}_{\perp}=x^{\mu}-x^{\mu}_{\parallel}$. $V(r)$ is a scalar
function of $r=z_{\perp}$.

  Choosing the model harmonic interaction,
$V(r)=(4/3)V_{0}r^{2}$, $V_{0}=const$, just as in ref. \cite{hkkp},
we are able to use the SDE and BSE solutions for the pion
and the kaon, which were obtained in \cite{hkkp}, and use them here in
the calculation of ${\bar K}^{0}\rightarrow\gamma\gamma$ .
The case of the ``funnel'' (Coulomb+linear) potential has also been
solved \cite{funnel}, but we do not use it here to avoid complexities
of the renormalization of the divergences appearing in the
bound-state equations \cite{AdDav,AlkAm,funnel} for this choice of
the potential. On the other hand, when $V(r)$ is chosen
to be the Nambu-Jona-Lasinio contact potential, a UV cut-off is needed.
An additional motivation for choosing the harmonic potential is
therefore the fact that both the SDE and the BSE are then divergence-free.
Namely, besides the absence of divergences in quark-loop integrals
commented on in the Introduction, we also avoid divergences in the
bound-state equations themselves, so that no regularizations or
cut-offs are necessary.

Since in the presently calculated matrix elements
we need only one bilocal, it is
of course most convenient to work in its rest frame.
We point out that
in this frame the special ansatz (\ref{eq:defk}) for the
interaction kernel reduces
$[\eta=(1,0,0,0)]$ to the ordinary $\gamma^{0}V(r)\gamma^{0}$ type of
interaction used in many calculations for mesons in the ``pairing"
(Nambu--Jona-Lasinio--inspired) approach, {\it e.g.} refs.
\cite{AdDav,ley,AlkAm}. Among these, we may point
out Le Yaouanc {\it et al.} \cite{ley} as a paradigmatic example
because they studied the harmonic binding, $V\sim r^{2}$, in detail.
However, the problems they pointed out as induced by non-covariance
(ambiguities in the definition of $F_{\pi}$, the wrong dispersion
relation)
are avoided by the usage of the ansatz kernel (\ref{eq:defk}) since
it leads to boost-invariant SDE and BSE. Of course, the important
features of pseudoscalar meson physics established in ref. \cite{ley},
such as the dynamical chiral symmetry breaking (D$\chi$SB) by
generating the dynamical quark mass and the appearance of the pion
as a Goldstone boson in the chiral limit, were also present in ref.
\cite{hkkp} and are therefore also present in this work.
This is especially
important here where we want to elucidate some aspects of
long-distance, non-perturbative QCD effects in
$\bar{K}^{0}\rightarrow\gamma\gamma$ .

The transitions of bilocal mesons are due to the interaction part
of the effective bilocal action $W$ used in ref. \cite{hkkp},
which we call $W_{trans}$.
It is obtained \cite{PeReEb,pervPL,perv} by integrating out the fermions
in the generating functional, which results in the fermion determinant,
{\it i. e.} a trace-log form whose expansion has infinitely many
terms:
\beq
    W_{trans}[{\cal M},L]=iN_{c}\sum_{n=2}^{\infty}\frac{1}{
    n}\mbox{Tr}\Phi^{n} \qquad ,
\label{eq:defw}
\eeq
where ``Tr'' also includes the integration, and $\Phi$ is defined
through the meson bilocal ${\cal M}$ which must now be
``shifted" \cite{hkkp,KaWe} by the transition-inducing
external operator $L$, coupled locally to the internal quark lines
represented by the dressed propagator $G_{\Sigma}$:
\beq
   \Phi(x,y)=\int d^{4}z G_{\Sigma}(x,z)[{\cal M}(z,y)-L(z)\delta^{
     (4)}(x-y)] \qquad .
\label{eq:defp}
\eeq
For the problem at hand, $L(x)$ represents a sum of operators describing
electromagnetic and weak radiative decays of mesons
(whose hadronic structure is described by the bilocal ${\cal M}$, the
solution of the bound-state equations governed
by the quark interaction kernel $K(x,y)$).

For leptonic weak decays (e. g. when calculating $F_{\pi}
$ or $F_{K}$ in ref. \cite{hkkp}),
$L(x)$ was simply the leptonic current coupled to the V--A quark current.
Here, in order
to induce $\bar{K}^{0}\rightarrow\gamma\gamma$ , $L(x)$ includes both
the ordinary flavour diagonal photonic coupling $eQ\gamma^{\mu}A_{\mu}(x)$
and the electroweak FC vertices contained  in (\ref{eq:lagg}):
\begin{eqnarray}
  V_{1\gamma}&=&B\varepsilon^{\mu\nu\lambda\rho}F_{\mu\nu}(x)
  i \stackrel{\leftrightarrow}{\partial}_{\lambda}(x)\gamma_{\rho}
  \,\frac{1}{2}(1-\gamma_{5})
  \qquad ; \hspace{0.4cm} \stackrel{\leftrightarrow}{\partial}=
  \stackrel{\rightarrow}{\partial}-\stackrel{\leftarrow}{\partial}
  \quad ,
\label{eq:v1} \\
  V_{2\gamma}&=&2 e_{D}B\varepsilon^{\mu\nu\lambda\rho}
   F_{\mu\nu}(x) A_{\lambda}(x) \gamma_{\rho}
  \,\frac{1}{2}(1-\gamma_{5}) \quad .
\label{eq:v2}
\end{eqnarray}
Here $Q=\frac{1}{3} \, $diag(2,--1,--1) and $e_{D}=eQ_{D}=-e/3$.

  The transition matrix element is thus
\beq
  {\cal A}_{\bar{K}^{0}\gamma\gamma}=\langle\gamma(k,\sigma),\gamma(
k',\sigma')|W_{trans}[{\cal M}-eQA\!\!\!/-V_{1\gamma}-V_{2\gamma}]|\bar{K
}^{0}(p)\rangle \quad ,
\eeq
where $p,k,k'$ are the kaon and photon momenta, respectively, and $\sigma ,
\sigma'$ are photon polarizations.

We recall that in ref. \cite{hkkp} the $\pi^{0}\rightarrow\gamma\gamma$
transition was caused by the cubic ($n=3$) term from $W_{trans}$,
because it contained subterms with one meson bilocal ${\cal M}$ and two
photon fields $A_{\mu}$, yielding the amplitude corresponding to the
triangle graph:
\beq
{\cal A}_{\pi^{0}\gamma\gamma}=\langle\gamma(k,\sigma),\gamma(
k',\sigma')|iN_{c}\mbox{Tr}({\cal M}G_{\Sigma}eQA\!\!\!/G_{\Sigma}eQA\!\!
\!/G_{\Sigma})|\pi^{0}(p)\rangle\quad.
\eeq
Similarly, for strangeness-changing transitions, the cubic ($n=3$) term
will contribute to  $\bar{K}^{0}\rightarrow\gamma\gamma$ through
subterms containing one meson bilocal ${\cal M}$, one pure electromagnetic
and one single-photon effective electroweak vertex $V_{1\gamma}$:
\beq
{\cal A}^{(1)}_{\bar{K}^{0}\gamma\gamma}=\langle\gamma\gamma
|iN_{c}\mbox{Tr}({\cal M}G_{\Sigma}V_{1\gamma}G_{\Sigma}eQ
A\!\!\!/G_{\Sigma}+ {\cal M}G_{\Sigma}eQA\!\!\!/
G_{\Sigma}V_{1\gamma}G_{\Sigma})|\bar{K}^{0}\rangle \qquad .
\label{eq:a1}
\eeq
This corresponds to the triangle graphs in fig. 1 and their
crossed counterparts.

On the other hand, the existence of the FC two-photon vertex leads to
the contribution to the total amplitude from the quadratic ($n=2$) term:
\beq
{\cal A}^{(2)}_{\bar{K}^{0}\gamma\gamma}=\langle\gamma(k,\sigma),\gamma(
k',\sigma')|iN_{c}\mbox{Tr}({\cal M}G_{\Sigma}V_{2\gamma}G_{\Sigma})
|\bar{K}^{0}(p)\rangle \qquad ,
\label{eq:a2}
\eeq
corresponding to the graph in fig. 2. Obviously, this
graph is in essence given by the kaon decay constant $F_{K}$ as
calculated in ref. \cite{hkkp}.

In order to keep the notation close to that of refs. \cite{ep93,ep94},
let us first relate the  $\bar{K}^{0}\rightarrow\gamma\gamma$ matrix
elements to the $C_{\bar{K}^{0}}$ coupling and after that consider the
physical $C_{K_{2}}$ in (\ref{eq:lagp}).

   Observing that the sum of the amplitudes (\ref{eq:a1}) and
(\ref{eq:a2})
should match the matrix element of the corresponding operator
(\ref{eq:lagp}), we can write
\begin{eqnarray}
   8\alpha C_{\bar{K}^{0}}& = & {\cal A}^{(2)}_{\bar{K}^{0}\gamma\gamma}
  -  {\cal A}^{(1)}_{\bar{K}^{0}\gamma\gamma} \nonumber \\
    & = & -4e_{D}B(F_{K}-F_{LD}) \qquad .
\label{eq:str}
\end{eqnarray}
Here the effect is expressed by the familiar decay constant $F_{K}$
and by the generalization of the pion-decay triangle loop amplitude
$F_{LD}$. This way of expressing the amplitude is
very transparent and enables one to
understand the effect both qualitatively and quantitatively.

The amplitudes of definite strangeness in (\ref{eq:str}) are then the
building blocks needed to construct the physical $K_{L}$-decay amplitude.
Restricting ourselves to the overwhelming CP-conserving amplitude from the
$K_{2}=\frac{1}{\sqrt{2}}(K^{0}+\bar{K}^{0})$, CP=--1 eigenstate,
we obtain the sought  decay strength as
\beq
  C_{K_{2}}=G_{F}\lambda_{u}\frac{\hat{B}_{u}}{
  6\pi}(F_{K}-F_{LD}) \qquad .
\label{eq:cp}
\eeq
The main discussion
in the concluding section is devoted to the amplitudes $C_{K_{2}}$ and
$C_{K_{2}}^{\sigma}$. Whereas $C_{K_{2}}$ results from the full
FC effective Lagrangian
${\cal L}_{\gamma}$ in (\ref{eq:lagg}),
$C_{K_{2}}^\sigma$ is the amplitude resulting just
from its on-shell remnant ${\cal L}_{\sigma}$ in (\ref{eq:lags}).

\section{Results and discussion}

We present our numerical results in two tables, where
 the potential strength $V_0$ and the current quark masses $m_q$
 are the only input model parameters. The other quantities are
model outputs, except ${\hat B}_u$, which lies within an
interval of values, depending on the scale parameter $\mu$
at which the operators in (\ref{eq:lagg})--(\ref{eq:lags})
are brought from
the original $M_W$ scale. Therefore we used two choices
which can be inferred from the discussion in the Introduction.
The choice (I) of using a common value (${\hat B}_u$ = 0.2)
as appropriate at higher $\mu$-values ($\sim 0.7$ GeV)
in fact maximizes the on-shell contribution.
The choice (II) of low values of $\mu$ (= 0.3 GeV)
implies larger but also significantly different ${\hat B}_u$'s
for different operators because of their different
anomalous dimensions, namely ${\hat B}_u = 0.66$ and
${\hat B}_u =0.32$ for ${\cal L}_\gamma$ and ${\cal L}_\sigma$,
respectively. This choice leads us closer to the empirical
$C_{K_{2}}$ amplitude, but in fact somewhat underestimates
the share of the on-shell amplitude.
In both tables
we choose the harmonic-potential strength $\frac{4}{3}V_{0}=$(289
MeV)$^{3}$, which reproduces the experimental pion and kaon masses
for the standard, well-established ratio $m_{s}/m_{u,d}$=25 \cite{don93}
(and for $m_{u,d}\approx 2$ MeV) and is
also close to the $V_{0}$ values used by other authors,
{\it e.g.} \cite{ley}.
Since the quantities pertinent for estimating the relevance
of off-shellness ($C_{K_{2}}$ and $C_{K_{2}}^\sigma$, as
clarified below) depend on $V_0$ in the same
way, their {\it relative} importance stays
the same for any other value of $V_0$. Thus, for our
purposes it suffices to tabulate the dependence on $m_q$.

  For easier comparison with the more familiar, famous
$\pi^{0}\rightarrow\gamma\gamma$ decay, let us first present the results
in table 1 in the $SU(3)$  limit. Such a symmetry limit of equal quark
masses ($m_{u}=m_{d}=m_{s}$) exhibits
(i)  identical solutions of SDE and BSE for $\pi^{0}$ and $\bar{K}^{0}$ ,
(ii) equal bound-state masses, $M_{\pi}=M_{K}$ and
(iii) includes the chiral limit, $m_{u,d,s}=0\quad.$

It is important to note that the axial-anomaly contribution
to the $\pi^{0}, \bar{K}^0 \rightarrow\gamma\gamma$ amplitudes
must be quark mass independent. Table 1 shows that
the amplitudes $C_{\pi}, C_{K_2}$ as calculated in our
bound-state model, satisfy this requirement very well
even relatively far from the strict chiral limit,
up to more than $m_q = (\frac{4}{3}V_{0})^{1/3}\times 0.1
\approx 30$ MeV. This important property also holds for
any other choice of the potential parameter $V_0$.
This testifies that the bound-state approach considered
correctly reproduces the mass-independent anomaly behaviour.
Such qualitatively correct behaviour of the amplitude is for us more
instructive than its absolute size.

Another important qualitative feature is that
for any potential strength $V_0$, the meson masses $M_\pi$ and $M_K$
calculated in the present bound-state model behave as
the masses of (pseudo-)Goldstone bosons must behave because
of PCAC, namely $M_{\pi,K} \propto {\sqrt m_q}$.
However, both tables show that, for the parameters which
fit $\pi$ and $K$ masses well, the decay constants
$F_\pi, F_K$, the radiative decay amplitude  $C_{\pi}$,
 and especially $C_{K_2}$, come out systematically too small.
Such suppression seems to be a systematic feature of this
and similar brands of ``pairing" bound-state models, {\it e.g.}
 \cite{AdDav,ley,hkkp,funnel,KaWe}.
One general mechanism contributing to such suppression
in the Bethe--Salpeter approach relatively to the naive quark-model
value has been offered by Koniuk and collaborators \cite{koni} on
account of the dissolution of the $q\bar{q}$ Fock-space component
into multi-pair Fock-space components in the strongly bound systems.

Of course, with another choice of $V_0$ and $m_q$,
one can achieve much better agreement for meson decays if one accepts
meson masses which are several times too high, as shown in \cite{hkkp}
on the example of the pion. However, as already remarked above,
the problems of accounting quantitatively for the hadronic
structure and decays within the bound-state approach are out
of scope of this paper. Our study here focuses on the qualitative
issue whether there are significant off-shell effects in
weak decays of hadrons, or not.

How do we establish the presence of significant off-shell effects?
If they were zero or negligible, one could
drop ${\cal L}_F$ (4), as is often done, and instead of the
complete effective
$s \rightarrow d \gamma$ lagrangian (\ref{eq:lagg}) and (\ref{eq:lagfs})
use just the
off-diagonal magnetic-moment term ${\cal L}_\sigma$ (\ref{eq:lags}), the
part which survives when quarks are put on shell. Nevertheless,
as ${\cal L}_\sigma$ vanishes in the chiral limit ($m_{u,d}=m_s=0$),
the nonvanishing amplitude in this limit clearly demonstrates
the importance of off-shell effects.  In the nomenclature
of refs. \cite{ep93,ep94} followed in eqs. (\ref{eq:str}) and
(\ref{eq:cp}) in this paper, off-shellness manifests itself
by non-cancellation of $F_{LD}$
and $F_{K}$ (stemming from fig. 1 and fig. 2, respectively).

How do things change for non-vanishing current quark masses, $m_q \ne 0$?
Then ${\cal L}_{\sigma}\ne$ 0, and the
measure of the importance of the off-shellness is provided by
the difference between
$|C_{K_{2}}|$ and $|C_{K_{2}}^{\sigma}|$, resulting from lagrangians
(\ref{eq:lagg}) and (\ref{eq:lags}), respectively.
   Table 1 shows that not too far from the chiral limit the
situation remains
essentially the same as in this limit: $|C_{K_{2}}^{\sigma}|$ rises
approximately linearly with $m_{u,d}=m_{s}$, but remains quite small in
comparison with the complete $|C_{K_{2}}|$, which is constant in an
excellent approximation, as emphasized above.

Since $|C_{K_{2}}^{\sigma}|$ is artificially
small for the current quark masses, which are more appropriate for
 the pion than for the kaon, let us continue studying
the off-shell effect for finite
and non-degenerate quark masses in table 2.
We choose the aforementioned
$m_{s}/m_{u,d}=25$ value,
but we have in fact found the results to be quite stable for the
allowed $m_{s}/m_{u,d}\in (20,30)$ interval.
(At present, however, there are numerical limitations
in finding the solutions of bound-state equations for
$m_q$ significantly larger than 100 MeV, unless the
model scale  $\frac{4}{3}V_{0}$ is correspondingly
increased over its chosen value.)
Here we again find that $|C_{K_{2}}^{\sigma}|$ constitutes
less than 10\% of the total contribution, so that the off-shell
 effect is overwhelming.
So, the simplification which has often been used and which keeps only
${\cal L}_{\sigma}$ from the very beginning, thereby neglecting
off-shell contributions, is not
always justified, and certainly not for pseudoscalar mesons.

As an aside, note the onset of the significant non-anomalous part of
$|C_{K_{2}}|$ in the third row of table 2. There,
$|C_{K_{2}}|$ is increased by 30\% over the first row,
which is still fully anomalous, being almost the same as
the mass-independent
$|C_{K_{2}}|$ of the ``small-mass" table 1. This is as expected, since for
the realistic kaon, with the rather heavyish strange quark,
the non-anomalous
part of the direct two-photon amplitude should be of the same order
of magnitude as the anomalous part, in counterdistinction to $\pi^0$.

To our knowledge, in the literature there is no direct (non-pole) $\chi$PT
term of order  ${\cal{O}}(p^{4})$ responsible for
$K_L\rightarrow \gamma\gamma$.
Thus, similarly as in ref. \cite{ep94}, our dominating amplitude
provides an entirely
new direct-decay piece of order ${\cal{O}}(p^{4})$ .  The suppressed
matrix element is of higher order, ${\cal{O}}(p^{6})$,
and belongs to the same
class as the ambiguous reducible pole diagrams. We stick to the {\em
direct} amplitudes, interesting in their own right -- in the study of
the CP violation, which
actually triggered \cite{ep93} the study of the off-shellness.
The similar results of two very different approaches (the chiral
quark model
 evaluation \cite{ep93,ep94} and the present one) give us confidence
that we have achieved a qualitative understanding of the matrix elements
 at hand.
A more quantitative result will probably be obtained if more
refined BS approaches are applied. ({\it E. g.}, ref. \cite{JaMu}
with the ansatz gluon propagator.)
However, we are sure that our qualitative conclusions about
the importance of
off-shell effects will not change because of such an
improved description of
the hadronic structure. Namely, it seems impossible to envision
such hadronic
``dressing'' which would enhance the ${\cal L}_{\sigma}$ contribution above
the off-shell contribution calculated in the present model to be an
order of magnitude larger.

Let us stress that in the meantime ``the anatomy'' of the off-shellness
has been demonstrated in detail on the example of the heavy-light $B$
meson \cite{epa} and a comparative study of $K, B \rightarrow \gamma
\gamma$ decay \cite{epb}. The off-shell part of the $B \rightarrow \gamma
\gamma$ amplitude that can not be transformed away has been displayed on
Fig. 2 of ref. \cite{epa}. However, in order to calculate the $B \rightarrow
\gamma \gamma$ amplitude we had to invoke some bound-state model,
different from the chiral quark model applicable for $K_{L} \rightarrow
\gamma \gamma$  decay. As one might expect, the off-shell effect for $B
\rightarrow 2 \gamma$ turns out to be suppressed by the ({\em binding
energy})/$m_{b}$, but is still numerically interesting.  Therefore one
would welcome further investigation of this effect in the approaches
able to account for the nonperturbative QCD effects  also in the heavy
light systems.

An obvious candidate would be the QCD-sum-rule approach ({\it e.g.} ref.
\cite{nari}).  The necessary ingredient in such an approach is the
study of the mixing of the appropriate operators under the
renormalization \cite{tarr}.
 For the purposes of such future considerations, it might
therefore be worth
recalling the (at first sight surprising) result of ref [22], that the
operator in  $ {\cal L}_{F}$ (4) has zero anomalous dimension, whereas
the corresponding magnetic type operator  in $ {\cal L}_{\sigma}$ (5)
is well known (Grinstein {\it et. al.} in ref. \cite{svz}) to have a
nonvanishing anomalous dimension.
The observation that the photonic part
of the covariant derivative in  $ {\cal L}_{F}$ (4)
leads to an operator proportional to the
current, can be a quick way to  conjecture (in agreement with \cite{enp90})
 that the anomalous dimensions of the respective
operator should be zero.
  This has been  confirmed by an explicit calculation leading to
 eq. (5) in ref. [22].
 This fact can facilitate the consideration of the
mixing of the operator in eq. (\ref{eq:lagf}) under renormalization
with other operators of the same dimension.

 We believe that the QCD-sum-rule study of the operator in
(\ref{eq:lagf}) might shed further light on the bound-state/off-shell
effects, and deserves more investigation.
In turn, the consideration of the off-shell effects seems
to be unavoidable in
any attempt of a precise evaluation of the kaon direct-decay amplitudes.

\vskip 0.8cm

\noindent {\bf Acknowledgment}

I.P. thanks J.O. Eeg for helpful discussions.
The authors acknowledge the support of the EU contract CI1*--CT91--0893
(HSMU), and thank R. Baier for the kind hospitality at
Physics Department
of the Bielefeld University, where the main part of this work
was done.

\newpage

\section*{Tables}
\begin{table}[h]
\begin{center}
\begin{tabular}{|l|l|l|l|l|l|}
\hline
 $m_{d}=m_{s}$&$M_{\pi}=M_{K}$&$F_{\pi}=F_{K}$&
 $|C_{\pi}|$&
 $|C_{K_{2}}|_{(I)}$&
 $|C_{K_{2}}^{\sigma}|_{(I)}$  \\
\hline \hline

   0  &   0 &  33 &  1.8  & 1.7 &  0.0      \\ \hline
 2.0  & 140 &  34 &  1.8  & 1.7 &  0.009  \\ \hline
28.9  & 479 &  47 &  2.1  & 1.7 &  0.14   \\  \hline
\multicolumn{6}{|c|}{$C_{\pi}^{exp}=4.3\times10^{-4}$,
   $C_{K_{2}}^{exp}=59\times10^{-12}$}
\\ \hline
\end{tabular}
\end{center}
\caption{Pion and kaon masses $M_{\pi,K}$,
decay constants $F_{\pi,K}$ (in MeV), the absolute value of the $\pi^{0}
\rightarrow\gamma\gamma$ decay amplitude $|C_{\pi}|$ (in $10^{-4}$
MeV$^{-1}$) and the $K_{2}\rightarrow\gamma\gamma$ decay amplitudes
$|C_{K_{2}}|$ and $|C_{K_{2}}^{\sigma}|$ (in $10^{-12}$ MeV$^{-1}$) are
given for various values of the $SU(3)$-symmetric current quark
masses ($m_{s}=m_{u,d}$).
Throughout the table, $4/3\,V_{0}=(289\mbox{MeV})^{3}$ and $\hat{B}_{u}=
0.2$.}
\end{table}
\begin{table}[h]
\begin{center}
\begin{tabular}{|l|l|l|l|l|l|l|l|}
\hline
  $m_{d}$&$m_{s}$&$M_{K}$&$F_{K}$&
 $|C_{K_{2}}|_{(I)}$&
 $|C_{K_{2}}^{\sigma}|_{(I)}$&
 $|C_{K_{2}}|_{(II)}$&
 $|C_{K_{2}}^{\sigma}|_{(II)}$  \\
\hline \hline

 2.4 &  61 & 500 & 47 & 1.8 & 0.16 & 5.9 & 0.25 \\ \hline
 3.5 &  87 & 577 & 53 & 2.0 & 0.25 & 6.7 & 0.39 \\ \hline
 4.6 & 116 & 646 & 60 & 2.5 & 0.38 & 8.1 & 0.61  \\ \hline
 \multicolumn{8}{|c|}{$C_{K_{2}}^{exp}=59\times10^{-12}$}
\\  \hline
\end{tabular}
\end{center}
\caption{The kaon mass $M_{K}$, the decay constant $F_{K}$ (in MeV),
and the absolute
values of the $K_{2}\rightarrow\gamma\gamma$ decay amplitudes
$|C_{K_{2}}|$ and $|C_{K_{2}}^{\sigma}|$ (in $10^{-12}$ MeV$^{-1}$) are
given for various values of the current quark masses $m_{u,d}$ and
$m_{s}$ (respecting the ratio $m_{s}/m_{u,d}=25$).
Two cases are shown: (I) with $\hat{B}_{u}=0.2$, and (II)~with
$\hat{B}_{u}=0.66$ in the calculation of $|C_{K_{2}}|$, but
$\hat{B}_{u}=0.32$
in the calculation of $|C_{K_{2}}^{\sigma}|$.
Throughout the table, $4/3\,V_{0}=(289\mbox{MeV})^{3}$.}
\end{table}

\section*{Figure captions}

\noindent
Figure 1: Decay of the kaon bilocal due to the
one-photon flavour-changing vertex $V_{1\gamma}$. Dressed quark
propagators emanating out of the bound-state vertex and circling
around the loop have dynamically generated mass. \\[1ex]
\noindent
Figure 2: Decay of the kaon bound state due to the two-photon
flavour-changing vertex $V_{2\gamma}$. As in figure 1, the internal
lines are dressed quark propagators.

\end{document}